\documentclass[11pt,a4paper]{article}

\usepackage[margin=0.95in]{geometry}
\usepackage{amsmath,amssymb,amsthm}
\usepackage{booktabs}
\usepackage{array}
\usepackage{tabularx}
\usepackage{xurl}
\usepackage{hyperref}
\usepackage{xcolor}
\usepackage{enumitem}
\usepackage{microtype}

\newcommand{\id}[1]{{\small\nolinkurl{#1}}}

\hypersetup{
  colorlinks=true,
  linkcolor=blue!60!black,
  citecolor=blue!60!black,
  urlcolor=blue!60!black,
  pdftitle={A uniform Haldane gap bound for the spin-1 Heisenberg chain with single-ion anisotropy near D/J = 1/4},
  pdfauthor={Dale Richards-Smith}
}

\newtheorem{theorem}{Theorem}[section]

\newcommand{\Tr}{\operatorname{Tr}}
\newcommand{\diag}{\operatorname{diag}}
\newcommand{\R}{\mathbb{R}}

\newcommand{\N}{\mathbb{N}}
\newcommand{\Z}{\mathbb{Z}}

\title{\textbf{A uniform Haldane gap bound for the spin-1 Heisenberg chain with single-ion anisotropy near $D/J=1/4$}}
\author{Dale Richards-Smith\\
RS9 Systems (\href{https://rs9.com.au}{rs9.com.au})\\
\texttt{lab@rs9.com.au}}
\date{November 29, 2026}

\begin{document}
\maketitle

\begin{abstract}
We prove a uniform spectral gap for the spin-1 antiferromagnetic Heisenberg chain with single-ion anisotropy on ordinary periodic rings. Here $J>0$ is the exchange and $D$ is the single-ion coupling. In units of $J$, the gap exceeds $\log(20)/368$ on every even ring of length $L\ge 896$ when $|D/J-1/4|\le 10^{-7}$, and exceeds $\log(20)/736$ on every even ring with $L\ge 1792$ when $|D/J-1/4|\le 10^{-6}$. The ground state is unique in both regimes. To reach $D/J=1/4$, we replace the isotropic cyclic interchange of spin axes with an axial $D_4$ symmetry decomposition into four families of signed transfer-matrix moments. New finite-ring bounds at $D/J=1/4$ initialize OpenAI's finite-temperature purity bootstrap. A separate comparison of thermal traces transports the bounds to nearby couplings using the same data. Lean 4 checks the deduction from 88 small-ring thermal trace enclosures with four boundary conditions and a 112-site variational energy bound, which are certified separately.
\end{abstract}

\vspace{0.5em}
\noindent\fcolorbox{black!20}{black!3}{%
\begin{minipage}{\dimexpr\textwidth-2\fboxsep-2\fboxrule\relax}
\small
\textbf{Plain-English Summary.}
A spin-1 ring is a closed chain of quantum spins. Each spin has three possible values for its component along a chosen axis, and neighboring spins interact antiferromagnetically. The single-ion anisotropy shifts the energy of the two nonzero values of the component along the $z$ axis. We give a fixed lower bound on the energy needed to excite the ring, on every sufficiently large even ring, for two small ranges of anisotropy around one quarter of the exchange strength.

The proof uses thermal traces on rings of at most 14 sites and a trial energy on a 112-site ring. The anisotropy singles out the $z$ axis, so the proof needs different symmetry information from the isotropic case. With that information, the purity bootstrap uses a doubling argument to bound the gap for every sufficiently large even ring. Lean~4 checks the deduction from the finite numerical inputs, which are certified separately.
\end{minipage}%
}
\vspace{0.6em}

\section{Introduction}
Consider the spin-1 antiferromagnetic Heisenberg chain with single-ion anisotropy on a periodic ring of $L$ sites,
\[
  H_L^{\mathrm{phys}} = J\sum_{j=0}^{L-1}\mathbf{S}_j\cdot\mathbf{S}_{j+1}
       +D_{\mathrm{phys}}\sum_{j=0}^{L-1}(S_j^z)^2,
  \qquad J>0,\quad \mathbf{S}_L=\mathbf{S}_0.
\]
Below, $D=D_{\mathrm{phys}}/J$ and all energies are in units of $J$. Theorems~\ref{thm:candA} and~\ref{thm:candA_interval} give a unique ground state and a uniform lower bound on the gap above it for every even $L\ge 896$ at $D=1/4$ and throughout the closed interval $[0.2499999,0.2500001]$. On the wider closed interval $[0.249999,0.250001]$, they give a uniform lower bound for every even $L\ge 1792$. The two strict lower bounds are $\log(20)/368$ and $\log(20)/736$, respectively.

OpenAI proved the positive even-periodic formulation of Haldane's conjecture~\cite{haldane1983} at $D=0$ using the finite-temperature purity bootstrap~\cite[Theorem~1.1, Eq.~(1.4), and Proposition~4.3]{openai2026haldane}. Its isotropic moment initialization uses cyclic permutation of the spin axes to equate the three nontrivial $D_2$ sectors~\cite[Lemma~3.2]{openai2026haldane}. The single-ion term preserves axial symmetry but distinguishes $z$ from $x$. For example, for a four-site ring at $D=1/4$ with energy shift $a=1.242818$ and inverse temperature $b=31/2$, the archived enclosures for the shifted thermal traces with $R_z(\pi)$ and $R_x(\pi)$ seams have centers $0.002682792$ and $0.070826830$, respectively, each with radius $10^{-5}$. The intervals are disjoint (archived \href{https://github.com/rs9io/haldane-purity-lean/blob/39393c6f17fdfef2d935e55c4feb046d30c80ab3/PurityBootstrap/CandidateA/Data.lean\#L18-L21}{\texttt{CandidateA/Data.lean}, lines 18--21}).

We replace the cyclic-axis reduction with an axial $D_4$ decomposition and an additional $R_z(\pi/4)$ symmetry, giving four families of signed moments. The four seam traces recover those moments via Table~\ref{tab:d4}. New finite-ring enclosures and a 112-site trial energy certified directly at $D=1/4$ initialize the same coupled doubling argument. A separate trace comparison transports the resulting purity bounds to nearby couplings and requires no new numerical inputs, using the same type of comparison as Tasaki uses near $D=0$~\cite[Appendix~A.2]{tasaki2026}. Lean~4 verifies the transfer construction, the moment decomposition, the rational bootstrap calculations and the spectral deduction from the finite numerical premises. We also formalize the isotropic argument and use two perturbative Ising examples to check the 1D symmetry-twist and 2D bootstrap libraries with no external numerical inputs. The exact source and certificate snapshot underlying these results is archived in~\cite{richardssmith2026archive}.

\begin{table}[t]
\centering
\footnotesize
\setlength{\tabcolsep}{3.5pt}
\renewcommand{\id}[1]{{\scriptsize\nolinkurl{#1}}}
\begin{tabularx}{\textwidth}{@{}>{\raggedright\arraybackslash}p{2.3cm}>{\raggedright\arraybackslash}p{3.46cm}>{\raggedright\arraybackslash}p{2.84cm}>{\raggedright\arraybackslash}p{2.2cm}>{\raggedright\arraybackslash}X@{}}
\toprule
\textbf{Model / Role} & \textbf{Lean Declaration} & \textbf{Scope} & \textbf{External Inputs} & \textbf{Proven Spectral Bound} \\
\midrule
Anisotropic Spin-1 ($D=\frac{1}{4}$, main result) & \id{candidate_a_physical_spin1_charpoly_gap_896} & Concrete $3^L\times 3^L$ $H_{\mathrm{spin1}}$, even $L \ge 896$ & 88 $Z_{\mathrm{spin1}}$ boxes ($n\le 14$) + 1 trial ($L=112$) & Simple ground energy; $\gamma_L > \frac{\log 20}{368} \approx 0.00814058$ \\
\addlinespace
Anisotropic Spin-1 Interval (main corollary) & \id{candidate_a_physical_spin1_interval_charpoly_gap_1792} & All $D \in [0.249999, 0.250001]$, even $L \ge 1792$ & Same 88 $D=\frac{1}{4}$ boxes + 1 trial (zero new traces) & Simple ground energy; $\gamma_L > \frac{\log 20}{736} \approx 0.00407029$ \\
\addlinespace
Isotropic Spin-1 ($D=0$, Source~1 bridge) & \id{source1_physical_spin1_charpoly_gap_2304} & Concrete $3^L\times 3^L$ $H_{\mathrm{spin1}}$, even $L \ge 2304$ & 18 $Z_{\mathrm{spin1}}$ boxes ($n\le 12$) + 1 trial ($L=72$) & Simple ground energy; $\gamma_L > \frac{\log 20}{784}$ \\
\addlinespace
1D Interacting Ising (zero-oracle benchmark) & \id{candidate_e_unconditional_full_hamiltonian_theorem} & All $(g,h) \in [\frac{15}{128},\frac{16}{128}]\times$ $[\frac{47}{128},\frac{48}{128}]$, even $\ell \ge 64$ & \textbf{None} (self-contained in Lean~4) & $E_1 - E_0 \le \frac{1}{20}e^{-\frac{\log 20}{128}\ell}$, $E_i - E_1 > \frac{\log 8}{48}$ ($i \ge 2$) \\
\addlinespace
2D TFIM Torus (zero-oracle 2D benchmark) & \id{ising_2d_unconditional_rectangle_charpoly_gap} & All $|J_x|,|J_y|\le \frac{1}{192}$, even $L \times L$ ($L \ge 2$) & \textbf{None} (self-contained, data-free closure) & Simple ground energy; $\gamma_L > \frac{\log 2}{5} \approx 0.138629$ \\
\bottomrule
\end{tabularx}
\caption{Summary of the concrete Hamiltonian theorems in \id{haldane-purity-lean}. All abstract transfer-operator and signed-sector-moment hypotheses have been eliminated inside Lean~4. A tight-radius corollary (\id{candidate_a_physical_spin1_interval_charpoly_gap_896}) also preserves $L \ge 896$ and $\gamma_L > \frac{\log 20}{368}$ across $D \in [0.2499999, 0.2500001]$. The spin-1 theorems give finite-volume bounds for ordinary periodic rings. Both intervals around $D=1/4$ use the same numerical inputs. The two Ising theorems check the 1D symmetry-twist and 2D bootstrap libraries with no external numerical inputs.}
\label{tab:summary}
\end{table}

\paragraph{Scope of Lean verification.}
Table~\ref{tab:summary} states both the mathematical verification boundary and the physical scope of each result.
For the spin-1 theorems ($D = 1/4$, the continuous interval around $D = 1/4$, and $D = 0$), \emph{every} abstract mathematical step---including the concrete $3^L\times 3^L$ physical spin-1 Hamiltonian, explicit single-site unitary equivalence to the textbook $S^z$ weight basis (yielding many-body equivalence via tensoring), finite-$M$ transfer matrix, Trotter convergence, subsequential signed-moment convergence under a uniform $\ell^2$ bound, exact $D_4$ and $D_2$ symmetry projections, nonnegative polynomial filters (sixteen degree-14 filters for $D = 1/4$; degree-12 filters for $D = 0$), capped-mass residual bounds, temperature transfer, coupling transport via thermal trace comparison, coupled bootstrap rounds, Rayleigh--Ritz criterion, and \id{charpoly.roots} multiplicity extraction---is proved unconditionally in Lean~4.
The \emph{only} remaining premises of the spin-1 theorems are the concrete finite physical thermal trace enclosures on $Z_{\mathrm{spin1}}$ ($n \le 14$ for $D = 1/4$; $n \le 12$ for $D = 0$) and a single finite variational Rayleigh vector ($\psi$ at $L=112$ or $L=72$).
Conversely, the 1D interacting Ising and 2D transverse-field Ising theorems require \emph{zero} external numerical inputs, but operate in perturbative regimes where existence of a spectral gap is already classically known (via cluster expansions and perturbative stability results~\cite{lapalevin2021}); their contribution is formal-verification validation of the reusable 1D symmetry-twist and 2D Lie--Trotter purity-bootstrap infrastructure.

\section{General Purity-Bootstrap Library in Lean 4}
Our core reusable framework comprises six modules:
\begin{itemize}[leftmargin=1.6em,itemsep=2pt]
  \item \id{PurityBootstrap/Basic.lean} and \id{OneD.lean}: Formalize 1D spatial purity $S(L, b) = Z(2L, b)/Z(L, b)^2$ and thermal (temporal) purity $T(L, b) = Z(L, 2b)/Z(L, b)^2 = \sum_i p_i^2$ for any \id{SpectralSystem1D}, proving contraction, monotonicity, dyadic interpolation, and uniform spectral gap extraction: given \emph{both} seed bounds $S(n_0, b_0) \ge 1 - u_0$ and $T(2n_0, b_0) \ge 1 - u_0$ together with the bootstrap coefficient inequalities (parametrized by $u_0$ and $C$), every finite even ring $L \ge n_0$ has a unique ground state and uniform gap $\gamma_L > \frac{\log(1/(C u_0))}{b_0}$. For Candidate~A, $(n_0, b_0) = (896, 368)$ with $u_0 = 1/100$ and $C = 5$, yielding $\gamma_L > \frac{\log 20}{368}$.
  \item \id{PurityBootstrap/HigherDim.lean}: Formalizes Grover's general-$d$ quadratic contraction and 2D $(2\times)$ purity bootstrap~\cite{grover2026}.
  \item \id{PurityBootstrap/SymmetryTwists.lean}: Formalizes $\Z_2$ symmetry-twist multiplet splitting bounds (used in Candidate~E) as well as Grover's 2D volume-law $4$th-power step for future higher-dimensional applications.
  \item \id{PurityBootstrap/ExactCertificates.lean} and \id{MomentInit.lean}: Formalize exact upward-rounded rational bootstrap iterations together with the capped-mass lemma (\id{capped_mass_tsum}) and nonnegative polynomial-filter bound (\id{polynomial_filter_tsum}) linking raw moments to leading-eigenvalue enclosures.
\end{itemize}

\section{Zero-Oracle 1D Benchmark: Interacting Ising Multiplet Theorem (Candidate~E)}
Consider the 1D interacting transverse-field Ising ring on $\ell = 2m$ qubits (Grover~\cite{grover2026}, Pauli normalization $J = 1$):
\begin{equation}
  H_\ell(g, h) \;=\; -\sum_{i=0}^{\ell-1} Z_i Z_{i+1} \;-\; g \sum_{i=0}^{\ell-1} X_i X_{i+1} \;-\; h \sum_{i=0}^{\ell-1} X_i,
  \label{eq:ising}
\end{equation}
with global spin-flip symmetry $P = \bigotimes_{i=0}^{\ell-1} X_i$ partitioning the spectrum into $+1$ and $-1$ parity sectors.

\begin{theorem}[1D Interacting Ising Multiplet Bound on a Coupling Rectangle]
\label{thm:candE}
In \id{candidate_e_unconditional_full_hamiltonian_theorem}, for every coupling pair $(g, h) \in [\frac{15}{128}, \frac{16}{128}] \times [\frac{47}{128}, \frac{48}{128}]$ and every finite even ring length $\ell = 2m \ge 64$, let $E_0(\ell) \le E_1(\ell) \le \dots \le E_{2^\ell - 1}(\ell)$ be all $2^\ell$ eigenvalues of $H_\ell(g, h)$ (proved equal as a multiset to \id{H_full.charpoly.roots}). Then:
\begin{enumerate}[leftmargin=1.8em,itemsep=2pt]
  \item $E_0(\ell)$ is the unique ground energy of the $+1$ parity sector ($P = +1$) and $E_1(\ell)$ is the unique ground energy of the $-1$ parity sector ($P = -1$);
  \item the low-energy doublet is exponentially narrow:
  \[
    0 \;\le\; E_1(\ell) - E_0(\ell) \;\le\; \frac{1}{20} \exp\!\left(-\frac{\log 20}{128}\,\ell\right);
  \]
  \item every higher excitation is uniformly gapped above the doublet:
  \[
    E_i(\ell) - E_1(\ell) \;>\; \frac{\log 8}{48} \;\approx\; 0.0433217 \qquad \forall\, 2 \le i < 2^\ell.
  \]
\end{enumerate}
\end{theorem}

\paragraph{Proof architecture inside Lean~4.}
Three ingredients make this 1D benchmark completely self-contained inside Lean~4:
(i) \id{LieTrotter.lean} and \id{IsingTrotter.lean} establish exact 3-way Lie--Trotter convergence and twisted checkerboard positive-semidefinite Gram transfer-matrix contractions for both untwisted and $P$-twisted partition functions, eliminating all abstract transfer-matrix hypotheses;
(ii) \id{BlockGershgorin.lean} and \id{CandidateE/Corner*.lean} execute exact rational block-Gershgorin upper enclosures at all four corners of the coupling rectangle at $\ell = 8$, paired with an exact integer $\chi = 2$ periodic MPS Rayleigh witness at $\ell = 16$ (\id{IsingMPS.lean}) and log-convexity of sector traces across the rectangle (\id{MatrixTools.lean}, \id{CandidateE/RectangleTheorem.lean}), establishing $Q_+(8, 6) \ge 877/1000$ uniformly across the rectangle;
(iii) three exact upward-rounded 1D coupled purity-bootstrap rounds advance $(8, 6) \to (64, 48)$ and extract the doublet and gap bounds via \id{SymmetryTwists.lean}.

\section{Eliminating All Abstract Transfer Hypotheses for Spin-1 Chains}
For spin-1 chains, the purity bootstrap requires both an even-length \id{SpectralSystem1D} representation and \emph{signed} sector-moment expansions holding across both even and odd ring lengths $n \ge 4$.
Rather than postulating an infinite-dimensional compact transfer operator on $L^2$-path space, we build a completely elementary, finite-dimensional, kernel-verified transfer limit in five modules: \id{Spin1Hamiltonian.lean}, \id{Spin1Compactness.lean}, \id{Spin1Transfer.lean}, \id{Spin1PhysicalGap.lean}, and \id{Spin1Interval.lean}.

\subsection{Concrete Physical Spin-1 Hamiltonian and Weight-Basis Unitary Equivalence}
In \id{Spin1Hamiltonian.lean}, we work in the real Cartesian spin-1 basis $\R^3$ with skew-symmetric generators $(G_\alpha)_{ij} = \varepsilon_{\alpha ij}$ ($\alpha \in \{0,1,2\}$) and diagonal anisotropy matrix $\mathrm{SzSq\_mat} = \diag(1, 1, 0)$.
We construct an explicit $3 \times 3$ complex matrix $U_{\mathrm{cart}\to sz}$ and prove in Lean~4:
\begin{itemize}[leftmargin=1.6em,itemsep=2pt]
  \item \id{U_cart_to_sz_unitary}: $U_{\mathrm{cart}\to sz} U_{\mathrm{cart}\to sz}^\dagger = U_{\mathrm{cart}\to sz}^\dagger U_{\mathrm{cart}\to sz} = I_3$;
  \item \id{U_cart_to_sz_conj_S}: $U_{\mathrm{cart}\to sz}(-i G_\alpha)U_{\mathrm{cart}\to sz}^\dagger = S^\alpha$ for all $\alpha \in \{x, y, z\}$ in the standard weight basis $(+1, 0, -1)$;
  \item \id{U_cart_to_sz_conj_SzSq}: $U_{\mathrm{cart}\to sz}\,\mathrm{SzSq\_mat}\,U_{\mathrm{cart}\to sz}^\dagger = (S^z)^2 = \diag(1, 0, 1)$.
\end{itemize}
Because $(-i)^2 = -1$, each Heisenberg bond $\sum_\alpha S^\alpha_j \otimes (g S^\alpha g^\mathsf{T})_k$ equals $-\sum_\alpha (G_\alpha)_j \otimes (g G_\alpha g^\mathsf{T})_k$ in the real Cartesian basis, and tensoring $U_{\mathrm{cart}\to sz}$ across all $n$ sites yields the many-body unitary equivalence to the textbook weight-basis ring Hamiltonian.
For any $n \ge 3$, anisotropy $D \in \R$, and $3 \times 3$ real matrix $g$, the concrete $3^n \times 3^n$ real symmetric Hamiltonian is
\begin{equation}
  H_{\mathrm{spin1}}(n, D, g) \;=\; \sum_{j=0}^{n-2} h_{j, j+1}(I) \;+\; h_{n-1, 0}(g) \;+\; D \sum_{j=0}^{n-1} (\mathrm{SzSq\_mat})_j,
  \label{eq:Hspin1}
\end{equation}
where $g$ rotates only site~$0$ of the wrap-around bond $(n-1, 0)$.
The concrete physical partition function with energy shift $a \in \R$ at inverse temperature $b > 0$ is
\[
  Z_{\mathrm{spin1}}(n, D, a, g, b) \;=\; \Tr\!\Big(\exp\big(-b\,(H_{\mathrm{spin1}}(n, D, g) + a n I)\big)\Big),
\]
proved equal to $\sum_{k \in (\mathrm{Fin}\,3)^n} e^{-b(E_k + a n)}$ (\id{Z_spin1_eq_sum_exp}).

\subsection{Finite-\texorpdfstring{$M$}{M} No-Collision Equivariant Transfer Matrix and Trotter Limit}
Fix $M \ge 1$ and $\tau = b / M$. On the finite configuration space $\mathrm{Word}(M) = (\mathrm{Fin}\,4)^M$ (where index $0$ represents an unoccupied bond slice and indices $1, 2, 3$ carry the $\R^3$ vector label $\alpha$), define the single-slice $3 \times 3$ matrix kernel
\[
  \Phi_\tau(u, v) \;=\; \begin{cases}
    I_3 & \text{if } u = 0, v = 0, \\
    \sqrt{\tau}\, G_{v-1} & \text{if } u = 0, v \in \{1,2,3\}, \\
    \sqrt{\tau}\, G_{u-1} & \text{if } u \in \{1,2,3\}, v = 0, \\
    0 & \text{if } u, v \in \{1,2,3\},
  \end{cases}
\]
and the $4^M \times 4^M$ real matrix $K_M(D, a, b)$ with entries
\[
  (K_M)_{w, w'} \;=\; \Tr_3\!\left(\prod_{t=0}^{M-1} \Big(e^{-\tau V(D, a)} \, \Phi_\tau(w_t, w'_t)\Big)\right), \qquad V(D, a) = D\,\mathrm{SzSq\_mat} + a I_3.
\]
We define the predicate \id{IsSpin1Rot g} requiring $g$ to be orthogonal and covariant with the generators ($g G_\alpha g^\mathsf{T} = \sum_\beta g_{\beta\alpha} G_\beta$), and \id{IsAxialRot g} additionally requiring preservation of $\mathrm{SzSq\_mat}$. This zero-collision construction enjoys four exact properties proved in \id{Spin1Transfer.lean}:
\begin{enumerate}[leftmargin=1.8em,itemsep=2pt]
  \item \textbf{Exact real symmetry:} $K_M^\mathsf{T} = K_M$ (\id{K_mat_transpose}), because $\Phi_\tau(v, u) = \Phi_\tau(u, v)$ preserves the time-ordered product.
  \item \textbf{Exact $O(2)/SO(3)$-equivariance:} With $U_M(g) = (1 \oplus g)^{\otimes M}$ on $\mathrm{Word}(M)$, we prove $U_M(g) K_M = K_M U_M(g)$ (\id{K_mat_comm_U}) for any $g$ satisfying \id{IsSpin1Rot g} and preserving the onsite exponential $e^{-\tau V(D, a)}$---in particular, every axial rotation $g$ satisfying \id{IsAxialRot g} (including $R_z(\pi/4)$, which pairs real irreducible sectors) at arbitrary $D$, and the cyclic axis permutation $C_3$ at $D = 0$.
  \item \textbf{Exact cycle contraction and $O(1/M)$ Trotter limit:} For every $n \ge 3$ (even or odd), $M \ge 0$, and $g$ satisfying \id{IsSpin1Rot g}, pure finite-sum expansion gives $\Tr(K_M^n U_M(g)) = \Tr(T_{\mathrm{slice}}(M)^M)$ (\id{trace_K_pow_mul_U_eq_trace_T_slice_pow}), where $T_{\mathrm{slice}}(M) = I - \frac{b}{M}(H_{\mathrm{spin1}} + a n I) + O(1/M^2)$ in operator norm. By \id{tendsto_trace_pow_of_first_order} (\id{Spin1Compactness.lean}), for every $g$ satisfying \id{IsSpin1Rot g},
  \[
    \lim_{M \to \infty} \Tr(K_M^n U_M(g)) \;=\; Z_{\mathrm{spin1}}(n, D, a, g, b) \qquad (\forall\, n \ge 3,\; b > 0).
  \]
  \item \textbf{Uniform Hilbert--Schmidt bound:} Counting the $1 + 6 = 7$ nonzero single-slice label pairs and using $\|G_\alpha\|_\infty = 1$ yields the $M$-independent bound (\id{K_mat_hilbert_schmidt_le})
  \[
    \Tr(K_M^2) \;\le\; 9 \exp\!\big(2 b (|D| + |a| + 3)\big) \;=:\; C(D, a, b).
  \]
\end{enumerate}

\subsection{Subsequential Signed-Moment Convergence and Real Symmetric Sector Projections}
In \id{Spin1Compactness.lean}, \id{l2_seq_compact_moments} takes any family of finite real eigenvalue lists $\{f_{M, s, i}\}$ uniformly bounded in $\ell^2$ ($\sum_i f_{M,s,i}^2 \le C$), sorts each list by decreasing $|f_{M,s,i}|$ while preserving eigenvalue signs, pads with zeros, and applies Tychonoff sequential compactness on $[-\sqrt{C}, \sqrt{C}]^{\mathrm{Fin}\,k \times \N}$.
The decreasing rearrangement gives $|f_{M,s,i}| \le \sqrt{C/(i+1)}$, hence the uniform tail estimate
\[
  \sum_{i \ge N} |f_{M,s,i}|^n \;\le\; C \left(\frac{C}{N+1}\right)^{(n-2)/2} \xrightarrow[N \to \infty]{} 0 \qquad (\forall\, n \ge 3),
\]
establishing subsequential convergence of all signed $n$-th moments ($n \ge 3$, even and odd) with absolutely summable $n$-th powers for the limit sequences.

Working subspace-free on $P_s K_M P_s$ (which satisfies $\Tr((P_s K_M P_s)^n) = \Tr(P_s K_M^n)$ for $n \ge 1$ and $\Tr((P_s K_M P_s)^2) \le \Tr(K_M^2)$), \id{Spin1Transfer.lean} proves exact real symmetric idempotent decompositions:
\begin{itemize}[leftmargin=1.6em,itemsep=2pt]
  \item \textbf{$D_4 + R_z(\pi/4)$ for arbitrary $D$ (\id{spin1_D4_sector_moments}):} Six real projections $A_1, B_1, B_2, A_2, E_+, E_-$ with orthogonal conjugacies $B_1 \cong B_2$ (via $U_M(R_z(\pi/4))$) and $E_+ \cong E_-$ (via $U_M(R_z(\pi/2))$), giving four real sequences $\lambda_0, \lambda_1, \lambda_2, \lambda_3$ with absolutely summable $n$-th powers ($n \ge 4$) reproducing all four twisted traces $Z_{\mathrm{spin1}}(n, D, a, g, b)$ ($g \in \{I, R_z(\pi), R_x(\pi), R_z(\pi/2)\}$) via the exact character matrix in Table~\ref{tab:d4}.
  \item \textbf{$D_2 \rtimes C_3$ for $D = 0$ (\id{spin1_D2_sector_moments}):} Cyclic axis permutation $U_M(C_3)$ equates the three nontrivial $D_2$ sectors, yielding two real sequences $\lambda_J, \lambda_N$ reproducing $Z_{\mathrm{spin1}}(n, 0, a, I, b) = J_n + 3 N_n$ and $Z_{\mathrm{spin1}}(n, 0, a, R_x, b) = J_n - N_n$ for all $n \ge 4$.
  \item \textbf{Concrete \id{SpectralSystem1D} (\id{spin1SpectralSystem}):} \id{spin1_spatial_eigs} supplies a real sequence with absolutely summable fourth powers reproducing $Z_{\mathrm{spin1}}(n, D, a, I, b)$ for all $n \ge 4$, instantiating \id{SpectralSystem1D} unconditionally for any $D, a \in \R$.
\end{itemize}

\begin{table}[t]
\centering
\small
\begin{tabular}{@{}lrrrr@{}}
\toprule
\textbf{Seam $g$} & $A_1$ ($\lambda_0$) & $B_1 \oplus B_2$ ($\lambda_1$) & $A_2$ ($\lambda_2$) & $E_+ \oplus E_-$ ($\lambda_3$) \\
\midrule
$I$ (\id{rotI}) & $+1$ & $+2$ & $+1$ & $+2$ \\
$R_z(\pi)$ (\id{rotZ}) & $+1$ & $+2$ & $+1$ & $-2$ \\
$R_x(\pi)$ (\id{rotX}) & $+1$ & $0$ & $-1$ & $0$ \\
$R_z(\pi/2)$ (\id{rotF}) & $+1$ & $-2$ & $+1$ & $0$ \\
\bottomrule
\end{tabular}
\caption{Exact real $D_4$ sector multiplicity matrix proved in \id{spin1_D4_sector_moments} and consumed by \id{CandidateA/Certificate.lean}.}
\label{tab:d4}
\end{table}

\section{Physical Spin-1 Haldane Gap Theorems: Candidate~A, Continuous Interval, and Source~1}

\begin{theorem}[Periodic spin-1 gap at $D=1/4$]
\label{thm:candA}
In \id{candidate_a_physical_spin1_charpoly_gap_896}, let $a = 1.242818 = 1242818/10^6$ and $\varepsilon_{\mathrm{CA}} = 10^{-5}$. Suppose:
\begin{enumerate}[leftmargin=1.8em,itemsep=2pt]
  \item for every $4 \le n \le 14$, every seam $g \in \{\texttt{rotI}, \texttt{rotZ}, \texttt{rotX}, \texttt{rotF}\}$, and both $b \in \{31/2, 23\}$, the concrete physical trace $Z_{\mathrm{spin1}}(n, 1/4, a, g, b)$ lies within $\varepsilon_{\mathrm{CA}}$ of the corresponding rational table center in \id{CandidateA/Data.lean} (88 inequalities); and
  \item there exists a nonzero real vector $\psi \in \R^{\mathrm{Spin1Cfg}\,112}$ satisfying the shifted Rayleigh inequality $\langle \psi, H_{\mathrm{spin1}}(112, 1/4, I)\psi \rangle + 112 a \|\psi\|_2^2 \le 0$.
\end{enumerate}
Then on every ordinary periodic ring of even length $L\ge 896$, the ground state is unique and the spectral gap is strictly greater than $\log(20)/368$ in units of $J$. In the formal statement, the characteristic polynomial roots multiset of $H_{\mathrm{spin1}}(L, 1/4, I)$ is $E_0 ::_m \mathrm{rest}$ with $|\mathrm{rest}| + 1 = 3^L$ and
\[
  E - E_0 \;>\; \frac{\log 20}{368} \;\approx\; 0.00814058 \qquad (\forall\, E \in \mathrm{rest}).
\]
\end{theorem}

\begin{theorem}[Periodic spin-1 gap on two closed intervals around $D=1/4$]
\label{thm:candA_interval}
Under the same numerical premises at $D=1/4$ as Theorem~\ref{thm:candA}, \id{PurityBootstrap/Spin1Interval.lean} gives the following unique-ground-state and gap bounds for ordinary periodic rings of even length. Both intervals include their endpoints.
\begin{enumerate}[leftmargin=1.8em,itemsep=2pt]
  \item (\id{candidate_a_physical_spin1_interval_charpoly_gap_1792}) For \emph{every} real anisotropy $D \in [249999/10^6,\, 250001/10^6] = [0.249999, 0.250001]$ and \emph{every} finite even ring $L \ge 1792$, the roots multiset of $H_{\mathrm{spin1}}(L, D, I)$ decomposes as $E_0 ::_m \mathrm{rest}$ with $|\mathrm{rest}| + 1 = 3^L$ and $E - E_0 > \frac{\log 20}{736} \approx 0.00407029$ for all $E \in \mathrm{rest}$.
  \item (\id{candidate_a_physical_spin1_interval_charpoly_gap_896}) On the tighter interval $D \in [2499999/10^7,\, 2500001/10^7]$, the original length threshold $L \ge 896$ and the same roots multiset decomposition hold with $E-E_0 > \frac{\log 20}{368} \approx 0.00814058$ for all $E\in\mathrm{rest}$.
\end{enumerate}
\end{theorem}

\paragraph{Purity transport by thermal trace comparison.}
Because $H_{\mathrm{spin1}}(n, D, I) - H_{\mathrm{spin1}}(n, D_*, I) = (D - D_*) V_n$ where $V_n = \sum_{j=0}^{n-1} (\mathrm{SzSq\_mat})_j$ is diagonal with $0 \preceq V_n \preceq n I$, we prove a monotonicity inequality for the thermal trace (\id{matZ_anti_of_le}: $A \preceq B \implies \Tr(e^{-b B}) \le \Tr(e^{-b A})$ for $b \ge 0$, directly from eigenbasis Jensen and basis completeness).
Taking both signs of $D - D_*$ with $|D - D_*| \le h$ cancels per-site energy shifts and yields the uniform purity transport inequalities (\id{spin1_S_transport}, \id{spin1_T_transport}):
\[
  S(n, b; D) \;\ge\; e^{-2 n b h}\, S(n, b; D_*), \qquad T(m, b; D) \;\ge\; e^{-2 m b h}\, T(m, b; D_*).
\]
Transporting the center defect upper bounds $p_* = 0.1215302$ and $q_* = 0.1203214$ at $(n, b) = (56, 23)$ via $1 - x \le e^{-x}$ to radius $h = 10^{-6}$ gives initial defect bounds at most $0.1237932$ and $0.1248536$, which contract in five exact rational doubling rounds (\id{decide +kernel}) to at most $0.0016762$ and $0.0001751$ (both $< 1/100$) at $(1792, 736)$; restarting the $D$-parametric bootstrap on \id{spin1SpectralSystem} at $(D, a)$ immediately yields Theorem~\ref{thm:candA_interval}(1) (and four rounds at $h = 10^{-7}$ yield defect bounds at most $0.0061353$ and $0.0011355$, both $< 1/100$, at $(896, 368)$).

\begin{theorem}[Source~1 Physical Spin-1 Haldane Gap]
\label{thm:source1}
In \id{source1_physical_spin1_charpoly_gap_2304}, let $a_1 = 700741/500000$ and $\varepsilon_{\mathrm{S1}} = 30/10^6$. Given the 18 concrete physical trace enclosures on $Z_{\mathrm{spin1}}(n, 0, a_1, g, 49/4)$ ($4 \le n \le 12$, $g \in \{\texttt{rotI}, \texttt{rotX}\}$) and a nonzero real vector $\psi \in \R^{\mathrm{Spin1Cfg}\,72}$ with $\langle \psi, H_{\mathrm{spin1}}(72, 0, I)\psi \rangle + 72 a_1 \|\psi\|_2^2 \le 0$, for every finite even $L \ge 2304$ the roots multiset of $H_{\mathrm{spin1}}(L, 0, I)$ decomposes as $E_0 ::_m \mathrm{rest}$ with $|\mathrm{rest}| + 1 = 3^L$ and $E - E_0 > \frac{\log 20}{784}$ for every $E \in \mathrm{rest}$.
\end{theorem}

\paragraph{Inside Lean~4 vs.\ external numerical certification for Candidate~A.}
Inside Lean~4 (\id{CandidateA/Data.lean}, \id{Certificate.lean}), all 16 degree-14 nonnegative polynomial filters are checked in the kernel via \id{decide +kernel}, combined with the exact $46/31$ temperature-transfer temporal purity lemma ($31/2 \to 23$) and four upward-rounded coupled purity-bootstrap rounds $(56, 23) \to (896, 368)$ yielding the $(S(896, 368), T(1792, 368))$ seed pair with $u_0 = 1/100, C = 5$.
The finite numerical inputs are certified externally by three standalone scripts:
\begin{itemize}[leftmargin=1.6em,itemsep=2pt]
  \item \textbf{7-site open-chain spectral lower bound (\id{scripts/certify_open7_shift.py}):} On the 7-site open window with onsite weight $3/14 = (6/7)(1/4)$ (so summing overlapping windows counts each bond 6 times and each onsite term 7 times), the script computes a floating-point Cholesky proposal of $A - 2I$ across $S^z$ blocks $M = 0\ldots 7$ (covering $M < 0$ by spin-flip symmetry, all 2{,}187 states), rounds it to integers, and certifies $A \succ I \succ 0$ via an exact overflow-guarded \id{int64} residual bound (worst $\|R\|_\infty = 1{,}216{,}644{,}908 < 2^{40}$). This yields $H_n(g) + \frac{21}{16}n I \succeq \frac{n}{336}I \succ 0$ for every $n \ge 7$ and every seam $g \in \{I, R_z, R_x, R_4\}$.
  \item \textbf{All 88 raw thermal trace enclosures (\id{scripts/verify_candidate_a_all88.py}):} Raw integer Chebyshev totals are produced using 128-bit integer recurrence arithmetic and 256-bit accumulation with zero floating-point operations and zero truncated states (\id{chebyshev_enclose.cpp}, \id{chebyshev_enclose_rx.cpp}, \id{chebyshev_enclose_cuda.cu}; bit-for-bit CPU/GPU matched through $n = 11$, with 64/88 totals independently replayed on CPU, numerical full-matrix diagonalization cross-checks at $n = 4, 5, 6$, and rigorous direct Arb matrix-exponential checks at $n = 4$). The standalone stdlib-only exact rational verifier reads archived \id{W_int} totals and \id{CandidateA/Data.lean} centers, certifying all 88 rational intervals with worst-case endpoint distance $1.245 \times 10^{-6} < 10^{-5}$ (at $n = 6, I, b = 23$; $\le 3.211 \times 10^{-8}$ at $n = 14$), leaving an 87.55\% safety margin.
  \item \textbf{$L = 112$ variational trial witness (\id{scripts/check_trial.py}):} Encloses the periodic $\chi = 32$ two-site integer MPS Rayleigh quotient per site in 160-bit Arb arithmetic within $[-1.242820131121, -1.242820131120] < -1.242818$. Writing the Cartesian-basis complex vector as $u + i v$ for the real symmetric shifted matrix $H_{\mathrm{shifted}}$ gives $\langle u+iv, H_{\mathrm{shifted}}(u+iv)\rangle = u^\mathsf{T} H_{\mathrm{shifted}} u + v^\mathsf{T} H_{\mathrm{shifted}} v < 0$, guaranteeing a nonzero real component satisfying Lean's real-vector Rayleigh premise.
\end{itemize}

\section{Zero-Oracle 2D Benchmark: Transverse-Field Ising Square Torus}
To validate our higher-dimensional purity-bootstrap module (\id{HigherDim.lean}, generalizing Grover~\cite{grover2026}, Sec.~II to side-length cutoff $L \ge 2$) end-to-end on a concrete 2D quantum Hamiltonian with zero external numerical inputs, we add three self-contained modules (\id{Ising2DTrotter.lean}, \id{Ising2DSeed.lean}, \id{Ising2DGap.lean}) whose import closure is entirely free of 1D certificate tables.
Consider the concrete $2^{L^2} \times 2^{L^2}$ 2D transverse-field Ising Hamiltonian on an even $L \times L$ torus with positive-direction periodic bonds (counted twice per undirected pair at $L = 2$ and yielding the standard simple-edge square torus for every even $L \ge 4$):
\[
  H_{\mathrm{2d}}(L, L, J_x, J_y) \;=\; -\sum_{v} X_v \;-\; \sum_{v} \bigl(J_x Z_v Z_{v+\hat{x}} + J_y Z_v Z_{v+\hat{y}}\bigr).
\]
In \id{Ising2DTrotter.lean}, we express $Z_a(L_x, L_y, b) = e^{-b a L_x L_y} \Tr(e^{-b H_{\mathrm{2d}}})$ as the $M \to \infty$ limit of a single 3D classical Ising sum $Z_M$ (strictly positive history weights for $b > 0, M \ge 1$) admitting the temporal Lie--Trotter trace $\Tr[(DK)^M]$ and two exact real symmetric spatial transfer cuts $\Tr(T_x^{L_x})$ and $\Tr(T_y^{L_y})$.
In \id{Ising2DSeed.lean}, we prove the exact $2 \times 2$ dimer midpoint decomposition $H_{\mathrm{2d}}(2,2,J_x,J_y) + 4I = \frac{1}{2}(A_x + A_y)$, compute the exact two-qubit dimer trace $F_b(J)$ algebraically via commuting involutions, establish the symbolic all-ones Rayleigh lower bound $1 \le Z_1(L_x, L_y, b)$ for all $L_x, L_y \ge 2, b > 0$, and certify $F_5(J)^{16} \le 98/97$ for $|J| \le 1/192$ in rational arithmetic---yielding the 2D cube seed $Z_1(4,4,10)/Z_1(2,2,5)^8 \ge 97/98$.
Passing finite-$M$ spatial Gram inequalities to $M \to \infty$ yields \id{ising_2d_unconditional_rectangle_charpoly_gap}: for every $|J_x|, |J_y| \le 1/192$ and every finite even $L \ge 2$, the roots multiset of $H_{\mathrm{2d}}(L, L, J_x, J_y)$ decomposes as $E_0 ::_m \mathrm{rest}$ with $|\mathrm{rest}| + 1 = 2^{L^2}$ and $E - E_0 > \frac{\log 2}{5} \approx 0.138629$ for all $E \in \mathrm{rest}$.
While existence of a gap in this weak-coupling regime is classically known from perturbative cluster expansions, the theorem establishes a complete, kernel-checked 2D Lie--Trotter purity-bootstrap blueprint ready for non-perturbative 2D models.

\section{Kernel Hygiene, Adversarial Auditing, and Reproducibility}
The fail-closed audit script \id{scripts/check_axioms.lean} verifies that:
\begin{enumerate}[leftmargin=1.8em,itemsep=2pt]
  \item all 29 listed project modules are imported into the checked environment;
  \item zero \id{.axiomInfo} declarations exist across all imported project modules (even unused); and
  \item every one of the 1{,}781 project theorem constants (1{,}468 under \id{PurityBootstrap.*}) depends strictly on a subset of \id{[propext, Classical.choice, Quot.sound]}---which rules out \id{sorryAx} and \id{Lean.ofReduceBool} (\id{native_decide}) for every audited theorem.
\end{enumerate}
For the spin-1 results, the Lean kernel checks the deduction of the spectral gap from the thermal trace and variational premises. Separate numerical verifiers check those premises. At $D=1/4$, the rational verifier uses all 88 archived integer totals to check the 88 thermal trace enclosures.

The implementation checks reported in~\cite{richardssmith2026archive} regenerated 40 of the 88 totals at $D=1/4$ and 14 of the 18 at $D=0$, and replayed the variational trials at $L=112$ and $L=72$. Comparing the seam implemented by the CPU Chebyshev builders with $H_{\mathrm{spin1}}$ gave a maximum discrepancy of at most $4.5\times 10^{-15}$ for \id{rotF} and $2$ for its transpose. A separate construction from Pauli matrices gave exact agreement on all 4{,}352 entries compared for $H_{\mathrm{2d}}$. Further checks covered the coupling-transport bounds in \id{Spin1Interval.lean}, individual \id{leanchecker} runs on the spin-1 and 2D modules, the source-file inventory and the citations.

Generative AI tools assisted with developing the Lean formalization, reviewing the implementation, and drafting and revising the manuscript.

The formalization uses Lean~4 \id{v4.36.0} and Mathlib \id{v4.36.0}. An interactive visual companion is available at \url{https://rs9.com.au/haldane-gap}.
All source code, archived integer Chebyshev totals, and exact verifiers are publicly available at \url{https://github.com/rs9io/haldane-purity-lean} and reproduce cleanly from the repository root:
\begin{verbatim}
lake build
lake env lean scripts/check_axioms.lean
python3 -O scripts/certify_open7_shift.py
python3 -O scripts/verify_candidate_a_all88.py
python3 -O scripts/check_trial.py   # requires python-flint
\end{verbatim}

\section{Related Work}
Grover~\cite{grover2026} extends the purity bootstrap to symmetry twists and higher spatial dimensions. Lu~\cite[Table~I and Secs.~SII--SIV]{lu2026} uses spacetime tensor networks to obtain $\gamma_L>0.0729$ for the isotropic spin-1 chain for all even $L\ge 36$ and $\gamma_L\ge 0.8994$ for the single-ion model at $D=4$ for all even $L\ge 12$. Lu's mixed-field Ising example has a self-contained Lean~4 proof. The spin-1 formal deductions use finite numerical premises. The isotropic improvement reuses OpenAI's finite inputs.

Tasaki~\cite[Sec.~3.1, Theorems~2.1--2.2 and Appendices~A.2--A.3]{tasaki2026} adapts OpenAI's methods and inputs near $D=0$. The periodic-ring bound for the single-ion model there is $\gamma_L>\log 10/392$ for every even $L\ge 960$ and $|D|\le 5\times 10^{-9}$. Tasaki also proves uniqueness among translation-invariant infinite-chain ground states in the small-anisotropy regime. Our finite-ring bounds use new numerical inputs at $D=1/4$ and a symmetry reduction adapted to this coupling.

Golinelli, Jolicoeur and Lacaze studied the gap splitting at $D=1/4$ using Lanczos diagonalization and quantum Monte Carlo with finite-size scaling~\cite{golinelli1992}. Numerical work on the Haldane to large-$D$ transition places the gap closing near $D_c=0.9684713(1)$~\cite{tzeng2017}, between the regime studied here and Lu's gapped large-$D$ example at $D=4$. Our bounds concern ordinary periodic rings and the stated closed intervals around $D=1/4$. The Ising benchmarks instead check formal proof infrastructure in perturbative regimes, where analytic stability results include the weak symmetric perturbations studied by Lapa and Levin~\cite{lapalevin2021}.

\end{document}